\documentclass[conference]{IEEEtran}
\IEEEoverridecommandlockouts

\usepackage{cite}
\usepackage{amsmath,amssymb,amsfonts}
\usepackage{algorithmic}
\usepackage{graphicx}
\usepackage{textcomp}
\usepackage{xcolor}
\usepackage{braket}
\usepackage{url}
\usepackage{array}
\usepackage{makecell}

\begin{document}

\title{Entanglement Fidelity in Standard Quantum Channels}

\author{
     \IEEEauthorblockN{
        Niccolò Zanieri and Marios Kountouris}
     \IEEEauthorblockA{Andalusian Institute of Data Science and Computational Intelligence (DaSCI)\\
     Department of Computer Science and Artificial Intelligence\\
     University of Granada, Spain}
     }

\maketitle

\begin{abstract}
Entanglement fidelity quantifies how well a quantum channel preserves the correlations between a transmitted system and an inaccessible reference system. We derive closed-form expressions for the entanglement fidelity associated with several standard quantum noise models, including the random Pauli-$X$, dephasing, depolarizing, Werner-Holevo, generalized Pauli (Weyl), and amplitude-damping channels. For each model, we express the entanglement fidelity in terms of a general input density operator $\rho$, using Schumacher's Kraus-operator approach, which provides a channel-agnostic recipe applicable to any completely positive trace-preserving (CPTP) map with a finite Kraus representation. We then specialize to a communication scenario in which the source emits a two-letter parametric alphabet, thereby making explicit the dependence of entanglement preservation on both channel and source parameters. The resulting expressions enable direct comparisons of channel performance and rankings for representative families of input states, including common qubit states.
\end{abstract}

\section{Introduction}
Quantum communication protocols encode information into quantum states that are transmitted through a physical medium. In quantum Shannon theory, a memoryless source can be modeled as emitting ``letters'' drawn from an ensemble of states $\{\ket{\phi_x},q(x)\}$ \cite{preskill_notes} with specified probabilities $q(x)$. Even when the source emits pure states, the transmitted signal is most naturally described operationally by a density operator, which captures classical uncertainty, imperfect state preparation, and possible correlations between the transmitted system and other degrees of freedom. In the single-qubit setting considered throughout this work, we write
\begin{equation*}
\rho =
\begin{bmatrix}
a & c \\
\overline{c} & b
\end{bmatrix},
\end{equation*}
where $a, b\in\mathbb{R}$ and $c\in\mathbb{C}$, with $a+b=1$ and positivity constraints ensuring that $\rho\succeq 0$.

Realistic transmission is inevitably noisy: the system carrying the letter interacts with uncontrolled degrees of freedom, which can be modeled by coupling to an environment or, alternatively, by considering correlations with an inaccessible reference system. Such interactions may leave some local features of the reduced state nearly unchanged while substantially degrading its correlations with the reference. This distinction is particularly important in quantum information processing, where performance is often determined not only by local state preservation, but also by the preservation of entanglement and other nonclassical correlations with external systems.

A standard figure of merit for capturing this requirement is the \emph{entanglement fidelity}, introduced by Schumacher \cite{schumacher1996sendingquantumentanglementnoisy}. For a given input state $\rho$ and channel $\mathcal{N}$, $F_e(\rho,\mathcal{N})$ quantifies how faithfully $\mathcal{N}$ preserves the joint system-reference state associated with a particular purification of $\rho$. Entanglement fidelity is directly connected to central topics in quantum information theory, including quantum error correction \cite{nielsen1996entanglementfidelityquantumerror}, entanglement transmission and channel capacities \cite{Barnum_2000,devetak2004capacityquantumchannelsimultaneous}, and rate-distortion formulations with correlation-sensitive distortion measures \cite{Datta_2013}. It is also closely related to the structure of widely used protocols, such as teleportation, superdense coding, and quantum key distribution, in which a channel acts on one subsystem while correlated systems remain untouched \cite{Teleportation,Superdense-Coding,BB84,E91}.

Motivated by these considerations, this paper provides explicit, reusable expressions for entanglement fidelity under several noise models of practical interest and uses them to study how source design affects entanglement preservation. Our contributions are as follows:
\begin{itemize}
    \item We present a channel-agnostic procedure for evaluating $F_e(\rho,\mathcal{N})$ using Schumacher's Kraus (operator-sum) representation of $\mathcal{N}$, thereby deriving expressions that depend only on $\rho$ and the Kraus operators and avoid the need to model the reference system explicitly.
    \item We derive closed-form formulas for $F_e(\rho,\mathcal{N})$ for several standard channels, including the random Pauli-$X$, dephasing, depolarizing, Werner-Holevo (qubit case), generalized Pauli (Weyl), and amplitude-damping channels.
    \item We apply these formulas to a communication setting in which the source emits a two-letter parametric alphabet, thereby making explicit the dependence of $F_e$ on both channel and source parameters. This enables direct comparisons and rankings of channels for relevant families of input states, including standard qubit states.
\end{itemize}

The remainder of the paper is organized as follows. Section~\ref{sec:prel} introduces the necessary preliminaries, including quantum channels and Kraus representations. Section~\ref{fe-gen-sec:Fe-computation} develops the entanglement-fidelity framework and derives closed-form expressions for the single-qubit channels studied here. Sections~\ref{sec:two_letter_optimization} and~\ref{fe-two-sec:analysis} apply these results to two-letter parametric sources and analyze the dependence of $F_e$ on channel and source parameters. Finally, Section~\ref{fe-two-sec:rank} compares and ranks the considered channels for representative source states.

\section{Entanglement Fidelity}
\label{sec:prel}
Let $\mathcal{H}$ be a finite-dimensional Hilbert space, and let $\mathcal{B}(\mathcal{H})$ denote the algebra of linear operators on $\mathcal{H}$.
A \emph{quantum channel} is a completely positive, trace-preserving (CPTP) linear map
$\mathcal{N}:\mathcal{B}(\mathcal{H})\to \mathcal{B}(\mathcal{H})$.
By the Kraus representation theorem, any CPTP map admits an operator-sum representation~\cite{Nielsen_Chuang_2010}
\begin{equation}
\mathcal{N}(X)=\sum_{i} K_i X K_i^\dagger,\qquad X \in \mathcal{B}(\mathcal{H}),
\label{eq:kraus}
\end{equation}
where $\{K_i\}_i \subset \mathcal{B}(\mathcal{H})$ are \emph{Kraus operators} satisfying the completeness relation
\begin{equation}
\sum_{i} K_i^\dagger K_i = I,
\label{eq:kraus_tp}
\end{equation}
which is equivalent to trace preservation. Conversely, any map of the form~\eqref{eq:kraus} that satisfies~\eqref{eq:kraus_tp} defines a quantum channel. The Kraus representation is not unique: different Kraus families related by an isometry represent the same channel.

\emph{Entanglement fidelity}, introduced by Schumacher~\cite{schumacher1996sendingquantumentanglementnoisy}, quantifies how well a quantum channel preserves the entanglement between a system and an inaccessible reference. It is a state-dependent figure of merit, depending on both the input density operator $\rho$ and the quantum operation (channel) $\mathcal{N}$ acting on it. Importantly, although its definition involves a reference system, $F_e(\rho,\mathcal{N})$ admits equivalent expressions that can be evaluated directly from $\rho$ and $\mathcal{N}$, without explicitly modeling the reference system.

Entanglement fidelity plays a central role in the analysis of quantum communication and quantum information processing, including the assessment of transmission noise, the design and evaluation of quantum error-correcting codes, and the characterization of protocols whose performance hinges on the preservation of quantum correlations.

Let $\rho\in \mathcal{B}(\mathcal{H})$ be an input state, that is, $\rho\succeq 0$ and $\mathrm{Tr}(\rho)=1$. Let $\mathcal{H}_R$ be a reference Hilbert space, and let $|\Psi\rangle\in \mathcal{H}_R\otimes \mathcal{H}$ be any purification of $\rho$, meaning that $\mathrm{Tr}_R\!\left(|\Psi\rangle\langle\Psi|\right)=\rho$, where $\mathrm{Tr}_R$ denotes the partial trace over the reference system. The \emph{entanglement fidelity} of $\mathcal{N}$ with respect to $\rho$ is defined by
\begin{equation}
F_e(\rho,\mathcal{N})
:=\langle \Psi|\,(\mathcal{I}_R\otimes \mathcal{N})\!\left(|\Psi\rangle\langle\Psi|\right)\,|\Psi\rangle,
\label{eq:ent_fid_def}
\end{equation}
where $\mathcal{I}_R$ is the identity channel on $\mathcal{B}(\mathcal{H}_R)$. The quantity $F_e(\rho,\mathcal{N})$ satisfies $0 \leq F_e(\rho,\mathcal{N}) \leq 1$ and is independent of the particular purification chosen.

A convenient expression for the entanglement fidelity, suitable for explicit evaluation, is obtained by combining~\eqref{eq:ent_fid_def} with a Kraus representation of the channel. Let $\mathcal{N}(X)=\sum_i A_i X A_i^\dagger$ be a Kraus decomposition of $\mathcal{N}$. Then the \emph{entanglement fidelity} is given by
\begin{equation}\label{fe-gen-eq:Fe}
F_e(\rho,\mathcal{N})
= \sum_i \bigl|\mathrm{Tr}(\rho A_i)\bigr|^2
= \sum_i \mathrm{Tr}(\rho A_i)\,\mathrm{Tr}(\rho A_i^\dagger).
\end{equation}

If all Kraus operators are Hermitian, that is, $A_i^\dagger=A_i$ for all $i$, then~\eqref{fe-gen-eq:Fe} reduces to
\begin{equation}
F_e(\rho,\mathcal{N})
= \sum_i \bigl(\mathrm{Tr}(\rho A_i)\bigr)^2
= \sum_i \bigl|\mathrm{Tr}(\rho A_i)\bigr|^2,
\end{equation}
since $\mathrm{Tr}(\rho A_i)\in\mathbb{R}$ in this case.

Intuitively, values of $F_e\approx 1$ indicate that the channel acts nearly as the identity on the relevant input state, preserving both $\rho$ and its correlations with the reference system. Conversely, when $F_e\ll 1$, the channel strongly degrades the system-reference entanglement, potentially even when local properties of the output state appear to be only weakly affected.

\section{Entanglement Fidelity of Standard Quantum Channels}\label{fe-gen-sec:Fe-computation}
In this section, we derive closed-form expressions for the \emph{entanglement fidelity} of several commonly used quantum channels. We consider the general setting in which an arbitrary single-qubit state (letter) is transmitted through a channel. The corresponding density operator can be written as
\begin{equation}\label{fe-gen-eq:rho}
\rho =
\begin{bmatrix}
a & c \\
\overline{c} & b
\end{bmatrix},
\end{equation}
where $a,b\in\mathbb{R}$, $c\in\mathbb{C}$, and $\overline{c}$ denotes the complex conjugate of $c$, with $a,b\ge 0$, $a+b=1$, and $|c|^2\le ab$ ensuring that $\rho$ is a valid state. To evaluate $F_e(\rho,\mathcal{N})$ for each channel $\mathcal{N}$, we first specify a Kraus (operator-sum) representation of the channel and then apply~\eqref{fe-gen-eq:Fe}.

Our results are organized into two groups. The first three channels are Pauli-based and parametrized by an error probability $u \in [0,1]$. The fourth channel is parameter-free. The fifth channel is specified by a parameter matrix $P$. Finally, the last channel is parametrized by $\gamma$, as is standard for the amplitude-damping channel.

\subsection{Pauli Channels}
\subsubsection{Random Pauli-$X$ Channel}\label{fe-gen-sec:pauli-X}
The single-qubit \emph{random Pauli-$X$} (bit-flip) channel with error probability $u=p_X\in[0,1]$ acts as
\begin{equation}
\mathcal{N}_{u}(\rho) = (1-u)\rho + u\,X\rho X,
\label{eq:pauliX_channel}
\end{equation}
where $X$ is the Pauli-$X$ operator. A Kraus (operator-sum) representation is given by the two nonzero Kraus operators,
\begin{equation}
A_I=\sqrt{1-u}\,I,\qquad A_X=\sqrt{u}\,X,
\label{eq:pauliX_kraus}
\end{equation}
so that $\mathcal{N}_{u}(\rho)=A_I\rho A_I^\dagger + A_X\rho A_X^\dagger$ and $A_I^\dagger A_I + A_X^\dagger A_X = I$.
Equivalently, this channel can be viewed as a special case of the general Pauli-channel form
\(
\mathcal{N}(\rho)=\displaystyle \sum_{P\in\{I,X,Y,Z\}} K_P\,\rho\,K_P^\dagger
\),
with $K_I=\sqrt{1-u}\,I$, $K_X=\sqrt{u}\,X$, and $K_Y=K_Z=0$.

The \emph{entanglement fidelity} of this channel for the state~\eqref{fe-gen-eq:rho} is given by
\begin{equation}
\begin{aligned}
F_e(\rho,\mathcal{N}_{u})
&= \left|\mathrm{Tr}\!\left(A_X\rho\right)\right|^2 + \left|\mathrm{Tr}\!\left(A_I\rho\right)\right|^2 \\
&= \left|\mathrm{Tr}\!\left(\sqrt{u}\,X\rho\right)\right|^2 + \left|\mathrm{Tr}\!\left(\sqrt{1-u}\,I\rho\right)\right|^2 \\
&= u\,\bigl|\mathrm{Tr}(X\rho)\bigr|^2 + (1-u)\,|\mathrm{Tr}(\rho)|^2 \\
&= u\,\bigl|\mathrm{Tr}(X\rho)\bigr|^2 + (1-u),
\end{aligned}
\label{eq:fe_pauliX}
\end{equation}
where we used the linearity of the trace and the fact that $\mathrm{Tr}(\rho)=1$. In particular, for the maximally mixed state $\rho=I/2$, one has $\mathrm{Tr}(X\rho)=0$, and therefore
\[
F_e\!\left(\tfrac{I}{2},\mathcal{N}_{u}\right)=1-u.
\]

Next, we compute the term appearing in~\eqref{eq:fe_pauliX}. Since
\begin{equation}
X\rho =
\begin{bmatrix}
0 & 1\\
1 & 0
\end{bmatrix}
\begin{bmatrix}
a & c\\
\overline{c} & b
\end{bmatrix}
=
\begin{bmatrix}
\overline{c} & b\\
a & c
\end{bmatrix},
\end{equation}
we obtain
\begin{equation}
\mathrm{Tr}(X\rho)=\overline{c}+c=2\,\Re(c).
\end{equation}
Substituting this expression into~\eqref{eq:fe_pauliX} yields
\begin{equation}\label{fe-gen-fe-pauli-x}
F_e(\rho,\mathcal{N}_u)
= (1-u) + u\,\bigl|2\,\Re(c)\bigr|^2
= 1 + u\left[4\bigl(\Re(c)\bigr)^2 - 1\right],
\end{equation}
where $\Re(c)$ denotes the real part of $c$.
\smallskip

\subsubsection{Dephasing Channel}\label{fe-gen-sec:dephasing}
The \emph{dephasing channel}, also known as phase-flip or random Pauli-$Z$ channel, with dephasing probability $u\in[0,1]$, is given by
\begin{equation}
\mathcal{N}_{u}(\rho) = (1-u)\rho + u\,Z\rho Z,
\label{eq:dephasing_channel}
\end{equation}
where $Z$ is the Pauli-$Z$ operator. A Kraus (operator-sum) representation is given by the two Kraus operators
\begin{equation}
K_0=\sqrt{1-u}\,I,\qquad K_1=\sqrt{u}\,Z,
\label{eq:dephasing_kraus}
\end{equation}
so that $\mathcal{N}_{u}(\rho)=K_0\rho K_0^\dagger+K_1\rho K_1^\dagger$.

The \emph{entanglement fidelity} of the dephasing channel is given by
\begin{equation}
\begin{aligned}
F_e(\rho,\mathcal{N}_u)
&= \left|\mathrm{Tr}(K_1\rho)\right|^2 + \left|\mathrm{Tr}(K_0\rho)\right|^2 \\
&= \left|\mathrm{Tr}\!\left(\sqrt{u}\,Z\rho\right)\right|^2 + \left|\mathrm{Tr}\!\left(\sqrt{1-u}\,I\rho\right)\right|^2 \\
&= u\,\bigl|\mathrm{Tr}(Z\rho)\bigr|^2 + (1-u),
\end{aligned}
\end{equation}
where we used the fact that $\mathrm{Tr}(\rho)=1$.

For the state~\eqref{fe-gen-eq:rho}, we compute
\begin{equation}
Z\rho
=
\begin{bmatrix}
1 & 0\\
0 & -1
\end{bmatrix}
\begin{bmatrix}
a & c\\
\overline{c} & b
\end{bmatrix}
=
\begin{bmatrix}
a & c\\
-\overline{c} & -b
\end{bmatrix},
\end{equation}
and hence
\begin{equation}
\mathrm{Tr}(Z\rho)=a-b.
\end{equation}
Substituting this expression into the formula for $F_e$ yields
\begin{equation}\label{fe-gen-fe-deph}
F_e(\rho,\mathcal{N}_u)
= (1-u) + u(a-b)^2
= 1 + u\left[(a-b)^2 - 1\right].
\end{equation}
\smallskip

\subsubsection{Depolarizing Channel}\label{fe-gen-sec:depolarizing}
The single-qubit \emph{depolarizing channel} with parameter $u\in[0,1]$ is defined by
\begin{equation}
\mathcal{N}_u(\rho) = (1-u)\rho + u\,\frac{I}{2}.
\label{eq:depol_def}
\end{equation}
An equivalent operator-sum representation is given by~\cite{Nielsen_Chuang_2010}
\begin{equation}
\mathcal{N}_u(\rho)
= \left(1-\frac{3u}{4}\right)\rho
+ \frac{u}{4}\left(X\rho X + Y\rho Y + Z\rho Z\right),
\label{eq:depol_kraus_form}
\end{equation}
which yields the Kraus operators
\begin{equation}
\left\{
\sqrt{1-\frac{3u}{4}}\,I,\;
\frac{\sqrt{u}}{2}\,X,\;
\frac{\sqrt{u}}{2}\,Y,\;
\frac{\sqrt{u}}{2}\,Z
\right\}.
\label{eq:depol_kraus_ops}
\end{equation}

In the previous two subsections, we computed $\mathrm{Tr}(X\rho)$ and $\mathrm{Tr}(Z\rho)$. It remains to compute $\mathrm{Tr}(Y\rho)$. We have
\begin{equation}
Y\rho
=
\begin{bmatrix}
0 & -i\\
i & 0
\end{bmatrix}
\begin{bmatrix}
a & c\\
\overline{c} & b
\end{bmatrix}
=
\begin{bmatrix}
-i\overline{c} & -ib\\
ia & ic
\end{bmatrix},
\end{equation}
and therefore
\begin{equation}
\mathrm{Tr}(Y\rho)=-i\overline{c}+ic=i(c-\overline{c})=-2\,\Im(c),
\end{equation}
where $\Im(c)$ denotes the imaginary part of $c$.

Therefore, the \emph{entanglement fidelity} of the depolarizing channel takes the form
\begin{eqnarray}\label{fe-gen-fe-depo}
F_e(\rho, \mathcal{N}_u) & = & \left(1 - \frac{3u}{4}\right)
+ \frac{u}{4}\bigl|\mathrm{Tr}(X\rho)\bigr|^2 \nonumber \\
& & \quad
+ \frac{u}{4}\bigl|\mathrm{Tr}(Y\rho)\bigr|^2
+ \frac{u}{4}\bigl|\mathrm{Tr}(Z\rho)\bigr|^2 \nonumber \\
& = & 1 + \frac{u}{4} \left[ 4(\Re(c))^2 + 4(\Im(c))^2 + (a-b)^2 - 3 \right] \nonumber \\
& = & 1 + u\left[|c|^2 + \frac{(a-b)^2}{4} - \frac{3}{4} \right].
\end{eqnarray}

\subsubsection{Werner--Holevo Channel}\label{fe-gen-sec:we}
In this subsection, we derive the entanglement fidelity for the ``pure'' Werner-Holevo channel. As shown in~\cite{Werner_2002}, for any dimension $d\ge 2$, the linear map $\Lambda_{\mathrm{WH}}:\mathcal{B}(\mathbb{C}^d)\to \mathcal{B}(\mathbb{C}^d)$ defined by
\begin{equation}
\Lambda_{\mathrm{WH}}(\rho)
= \frac{1}{d-1}\bigl(\mathrm{Tr}(\rho)\,I - \rho^{T}\bigr)
\label{eq:werner_holevo_channel}
\end{equation}
is a valid quantum channel (CPTP). Here, $(\cdot)^{T}$ denotes the matrix transpose with respect to a fixed orthonormal basis.

For $d\ge 2$, the Werner--Holevo channel is often discussed as a canonical example of a channel that cannot transmit quantum information in the usual sense (for example, because it has vanishing quantum capacity in entanglement-breaking regimes). In the qubit case $d=2$, however, the map admits a particularly simple form. Since $\mathrm{Tr}(\rho)=1$, we have
\begin{equation}\label{fe-gen-wh-channel}
\Lambda_{\mathrm{WH}}(\rho)
= I - \rho^{T}
= Y\,\rho\,Y,
\end{equation}
where $Y$ is the Pauli-$Y$ operator. Consequently, a Kraus representation is given by the single Kraus operator $A_1=Y$, and the \emph{entanglement fidelity} is given by
\begin{equation}
F_e(\rho,\Lambda_{\mathrm{WH}})
= \bigl|\mathrm{Tr}(Y\rho)\bigr|^2
= \bigl|-2\,\Im(c)\bigr|^2
= 4\bigl(\Im(c)\bigr)^2,
\end{equation}
where $c$ is the off-diagonal entry of $\rho$, as in~\eqref{fe-gen-eq:rho}.
\smallskip
\subsubsection{Generalized Pauli (Weyl) Channel}\label{fe-gen-sec:weyl}
We next consider the generalized Pauli (Weyl) channel, whose capacity properties have been studied in~\cite{Datta-channels}. This channel is defined on a $d$-dimensional Hilbert space and is specified by a $d\times d$ probability matrix $P=(p_{ij})_{i,j=0}^{d-1}$ satisfying $\sum_{i,j} p_{ij}=1$.

In the qubit case ($d=2$), a convenient Kraus representation is given by
\begin{equation}
\left\{
\sqrt{p_{00}}\,I,\;
\sqrt{p_{10}}\,X,\;
\sqrt{p_{01}}\,Z,\;
\sqrt{p_{11}}\,XZ
\right\},
\label{eq:weyl_kraus_qubit}
\end{equation}
where $p_{00},p_{10},p_{01},p_{11}\ge 0$ and $p_{00}+p_{10}+p_{01}+p_{11}=1$.

With an appropriate choice of $P$, the channels of Sections~\ref{fe-gen-sec:pauli-X}, \ref{fe-gen-sec:dephasing}, \ref{fe-gen-sec:depolarizing} and~\ref{fe-gen-sec:we} can be recovered as special cases. Using the Kraus expression for $F_e$, together with the trace identities computed above, the entanglement fidelity of the generalized Pauli (Weyl) channel is given by
\begin{equation}
\begin{aligned}
F_e(\rho,\mathcal{N})
&\quad = p_{00}\,|\mathrm{Tr}(I\rho)|^2
+ p_{10}\,|\mathrm{Tr}(X\rho)|^2 \\
&\quad + p_{01}\,|\mathrm{Tr}(Z\rho)|^2
+ p_{11}\,|\mathrm{Tr}(XZ\,\rho)|^2 .
\end{aligned}
\label{eq:fe_weyl_general}
\end{equation}

To evaluate the remaining term, note that $XZ = -iY$. It follows that $\mathrm{Tr}(XZ\,\rho) = -i\,\mathrm{Tr}(Y\rho),
\qquad\Rightarrow\qquad
\bigl|\mathrm{Tr}(XZ\,\rho)\bigr|^2=\bigl|\mathrm{Tr}(Y\rho)\bigr|^2$.

Substituting $\mathrm{Tr}(\rho)=1$, $\mathrm{Tr}(X\rho)=2\Re(c)$, $\mathrm{Tr}(Y\rho)=-2\Im(c)$, and $\mathrm{Tr}(Z\rho)=a-b$ into the expression for $F_e$ yields
\begin{equation}
\begin{aligned}
F_e(\rho,\mathcal{N})
&= p_{00}
+ 4p_{10}\bigl(\Re(c)\bigr)^2
+ p_{01}(a-b)^2
+ 4p_{11}\bigl(\Im(c)\bigr)^2.
\end{aligned}
\end{equation}
\smallskip

\subsubsection{Amplitude-Damping Channel}\label{fe-gen-sec:ampdamp}
The amplitude-damping channel models energy-relaxation processes such as spontaneous emission. A standard Kraus representation is given by
\begin{equation}
A_0=\sqrt{\gamma}\,\ket{0}\!\bra{1},
\qquad
A_1=\ket{0}\!\bra{0}+\sqrt{1-\gamma}\,\ket{1}\!\bra{1},
\label{eq:ad_kraus}
\end{equation}
where $\gamma\in[0,1]$ is the damping probability. These operators satisfy $A_0^\dagger A_0 + A_1^\dagger A_1 = I$.

In the computational basis, the corresponding Kraus operators take the matrix form
\begin{equation}
A_0 =
\begin{bmatrix}
0 & \sqrt{\gamma}\\
0 & 0
\end{bmatrix},
\qquad
A_1 =
\begin{bmatrix}
1 & 0\\
0 & \sqrt{1-\gamma}
\end{bmatrix}.
\end{equation}

Using~\eqref{fe-gen-eq:Fe}, the \emph{entanglement fidelity} of the amplitude-damping channel for the state~\eqref{fe-gen-eq:rho} is given by
\begin{equation}
\begin{aligned}
F_e(\rho,\mathcal{N})
&= \bigl|\mathrm{Tr}(A_0\rho)\bigr|^2 + \bigl|\mathrm{Tr}(A_1\rho)\bigr|^2 \\
&= \mathrm{Tr}(A_0\rho)\,\mathrm{Tr}(A_0^\dagger\rho)
+ \mathrm{Tr}(A_1\rho)\,\mathrm{Tr}(A_1^\dagger\rho).
\end{aligned}
\end{equation}

We first compute the traces appearing in this expression:
\begin{equation}
\begin{aligned}
\mathrm{Tr}(A_0\rho)
&= \mathrm{Tr}\!\left(
\begin{bmatrix}
\overline{c}\,\sqrt{\gamma} & b\,\sqrt{\gamma}\\
0 & 0
\end{bmatrix}
\right) \\
&= \overline{c}\,\sqrt{\gamma},
\end{aligned}
\end{equation}

\begin{equation}
\begin{aligned}
\mathrm{Tr}(A_0^\dagger \rho)
&= \mathrm{Tr}\!\left(
\begin{bmatrix}
0 & 0 \\
a\,\sqrt{\gamma} & c\,\sqrt{\gamma}
\end{bmatrix}
\right) \\
&= c\,\sqrt{\gamma},
\end{aligned}
\end{equation}

\begin{equation}
\begin{aligned}
\mathrm{Tr}(A_1\rho)
&= \mathrm{Tr}\!\left(
\begin{bmatrix}
a & c \\
\overline{c}\,\sqrt{1-\gamma} & b\,\sqrt{1-\gamma}
\end{bmatrix}
\right) \\
&= a + b\,\sqrt{1-\gamma}.
\end{aligned}
\end{equation}

Substituting these expressions into the formula for $F_e$, we obtain
\begin{equation}\label{fe-gen-fe-amp}
F_e(\rho,\mathcal{N})
= \gamma\,|c|^2 + \left(a + b\sqrt{1-\gamma}\right)^2.
\end{equation}
\smallskip

\subsection{Summary of the Results}
For ease of reference, we summarize the results in Table~\ref{fe-gen-tab:channels-fe}, where $\rho$ is defined as in~\eqref{fe-gen-eq:rho}.
\begin{table}[!tbh]
    \caption{Entanglement fidelities for the studied quantum channels}
    \centering
    \renewcommand{\arraystretch}{2.0}
    \begin{tabular}{|m{2.3cm}|m{5.6cm}|}
    \hline
    \multicolumn{1}{|c|}{\textbf{Channel $\mathcal{N}$}} & \multicolumn{1}{c|}{\textbf{$F_e(\rho,\mathcal{N})$}} \\ \hline
    Random Pauli-$X$ & $1 + u\left[4(\Re(c))^2 - 1\right]$ \\ \hline
    Dephasing & $1 + u\left[(a-b)^2 - 1\right]$ \\ \hline
    Depolarizing & $1 + \frac{u}{4}\left[4(\Re(c))^2 + 4(\Im(c))^2 + (a-b)^2 - 3\right]$\\ \hline
    Generalized Pauli (Weyl) & $p_{00} + 4p_{10}(\Re(c))^2 \newline + p_{01}(a-b)^2 + 4p_{11}(\Im(c))^2$ \\ \hline
    Amplitude-damping & $\gamma|c|^2 + \left(a + b\sqrt{1-\gamma}\right)^2$ \\ \hline
    Werner-Holevo & $4(\Im(c))^2$ \\ \hline
    \end{tabular}
    \label{fe-gen-tab:channels-fe}
\end{table}

\section{Optimization of a Two-Letter Message}
\label{sec:two_letter_optimization}
We now apply the previous results to a communication scenario in which the sender encodes a message using one of two possible quantum letters. Our objective is to characterize how the choice of the letter pair affects the preservation of entanglement between the transmitted system and an inaccessible reference. In particular, we show that an appropriate choice of letter pair can optimize the entanglement fidelity of the overall transmission.

Consider the following pair of quantum letters:
\begin{equation}\label{fe-two-eq:letters}
\begin{aligned}
\ket{\psi^+} &= \sqrt{p},\ket{0} + \sqrt{1-p},\ket{1},\\
\ket{\psi^-} &= \sqrt{p},\ket{0} - \sqrt{1-p},\ket{1}.
\end{aligned}
\end{equation}

The following density operator describes a system prepared in the state $\ket{\psi^+}$ with probability $q$ and in the state $\ket{\psi^-}$ with probability $1-q$:
\begin{equation}\label{fe-two-rho}
\rho = q \ket{\psi^+}\bra{\psi^+} + (1-q) \ket{\psi^-}\bra{\psi^-}.
\end{equation}

The density operator can be written explicitly as
\begin{equation}\label{fe-two-rho-explicit}
    \rho = \begin{bmatrix}
        p & (2q - 1) \sqrt{p(1-p)} \\
        (2q - 1) \sqrt{p(1-p)} & 1-p
        \end{bmatrix}.
\end{equation}

Depending on the values of $p$ and $q$, we obtain different source messages. There are five extreme cases, which are not representative of practical scenarios and lead to special behavior:
\begin{itemize}
    \item If $p = 0$ or $p = 1$, then, regardless of the value of $q$, the source reduces to a single-letter alphabet and therefore cannot convey any information.
    \item The cases $q = 0$ and $q = 1$ are analogous: the source again collapses to a single letter.
    \item If $p = \frac{1}{2}$, the two letters $\ket{\psi^+}$ and $\ket{\psi^-}$ are orthogonal, and the message can be treated as effectively classical (see, for example, Preskill~\cite{preskill_notes}).
\end{itemize}
The remaining parameter regimes correspond to the practically relevant case of messages drawn from a two-letter alphabet of \emph{nonorthogonal} quantum states, with both letters occurring with nonzero probability.

\subsection{Entanglement Fidelity Calculation}
For the state in~\eqref{fe-two-rho-explicit}, the parameters $a$, $b$, and $c$ defined in~\eqref{fe-gen-eq:rho} are real and take the following values:
\begin{equation}
\begin{aligned}
a &= p,\\
b &= 1-p,\\
c &= (2q-1)\sqrt{p(1-p)}.
\end{aligned}
\end{equation}

Using the results from Section~\ref{fe-gen-sec:Fe-computation}, we compute the \emph{entanglement fidelity} for several channels when the source message is built from the two letters in~\eqref{fe-two-eq:letters}. The resulting expressions are summarized in Table~\ref{fe-two-tab:channels-fe}.

\begin{table}[!tbh]
    \caption{Entanglement fidelities for the studied quantum channels, with $\rho$ defined as in~\eqref{fe-two-rho-explicit}.}
    \centering
    \renewcommand{\arraystretch}{2.5}
    \begin{tabular}{|m{2.5cm}|m{5.3cm}|}
    \hline
    \multicolumn{1}{|c|}{\textbf{Channel $\mathcal{N}$}} & \multicolumn{1}{c|}{\textbf{$F_e(\rho,\mathcal{N})$}} \\ \hline
    Random Pauli-$X$ & $1 + u\left[4(2q - 1)^2 p(1-p) - 1\right]$ \\ \hline
    Dephasing & $1 + 4u\,p(p-1)$ \\ \hline
    Depolarizing & \makecell{$1 + u\left[(2q - 1)^2 p(1-p) + \frac{(2p-1)^2}{4} - \frac{3}{4}\right]$} \\ \hline
    Generalized Pauli (Weyl) & \makecell{$p_{00} + 4p_{10}(2q-1)^2 p(1-p)$ \\ $+\, p_{01}(2p-1)^2$} \\ \hline
    Amplitude-damping & \makecell{$p^2 + (1-\gamma)(1-p)^2$ \\ $+\, p(1-p)\left[2\sqrt{1-\gamma} + \gamma(2q-1)^2\right]$} \\ \hline
    Werner-Holevo & \hspace{23mm} $0$ \\ \hline
    \end{tabular}    
    \label{fe-two-tab:channels-fe}
\end{table}

\section{Entanglement Fidelity Analysis}\label{fe-two-sec:analysis}
In general, the \emph{entanglement fidelity} of the channels considered here depends on three parameters: one parameter characterizing the channel (e.g., $u$ or $\gamma$) and two parameters specifying the source letters, namely $p$ and $q$.  
In this section, we present plots of the entanglement fidelities as functions of $p$ for representative values of $q$ and the channel parameter ($u$ or $\gamma$).

In a realistic scenario, not all of these parameters may be freely chosen. In particular, the parameter characterizing the channel may be fixed by the problem setting and thus lie beyond our control. Likewise, the value of $q$ is determined by the statistics of the source messages and may not be directly adjustable.

In contrast, the parameter $p$ specifies the particular quantum states used as letters and, in principle, can be selected by the sender through state preparation. While practical implementations may impose constraints on the admissible values of $p$, it still provides a degree of freedom in alphabet design. By exploiting this freedom, one can choose the pair of letters that maximizes performance, as quantified here by the entanglement fidelity, for the given channel conditions.

\subsection{Pauli channels}
All Pauli channels considered in this work are parametrized by $u$, which represents the error probability or noise strength. For $u = 0$, the channel reduces to the identity map and leaves the state unchanged. At the other extreme, $u = 1$ corresponds to the full application of the associated Pauli error or, depending on the channel, to a complete mixture over Pauli errors. Accordingly, we generally expect the entanglement fidelity to be higher for smaller values of $u$, since the channel acts more nearly as the identity and therefore preserves a larger fraction of the input state.

\subsubsection{Random Pauli-$X$ Channel}\label{fe-two-sec:pauli-x-an}
Recall from Table~\ref{fe-two-tab:channels-fe} that the entanglement fidelity for this channel is
\begin{equation*}
F_e(\rho, \mathcal{N}_u) = 1 + u \left[4(2q - 1)^2 p(1-p) - 1\right].
\end{equation*}
Since $F_e \in [0,1]$, the second term in the expression must be non-positive. Moreover, as anticipated, smaller values of $u$ lead to better preservation of the input state. This behavior is reflected in the fact that $F_e$ can have a strictly positive lower bound for sufficiently small $u$, meaning that the channel may be unable to destroy the entanglement completely.

\begin{figure}[ht!]
    \includegraphics[width=0.5\textwidth]{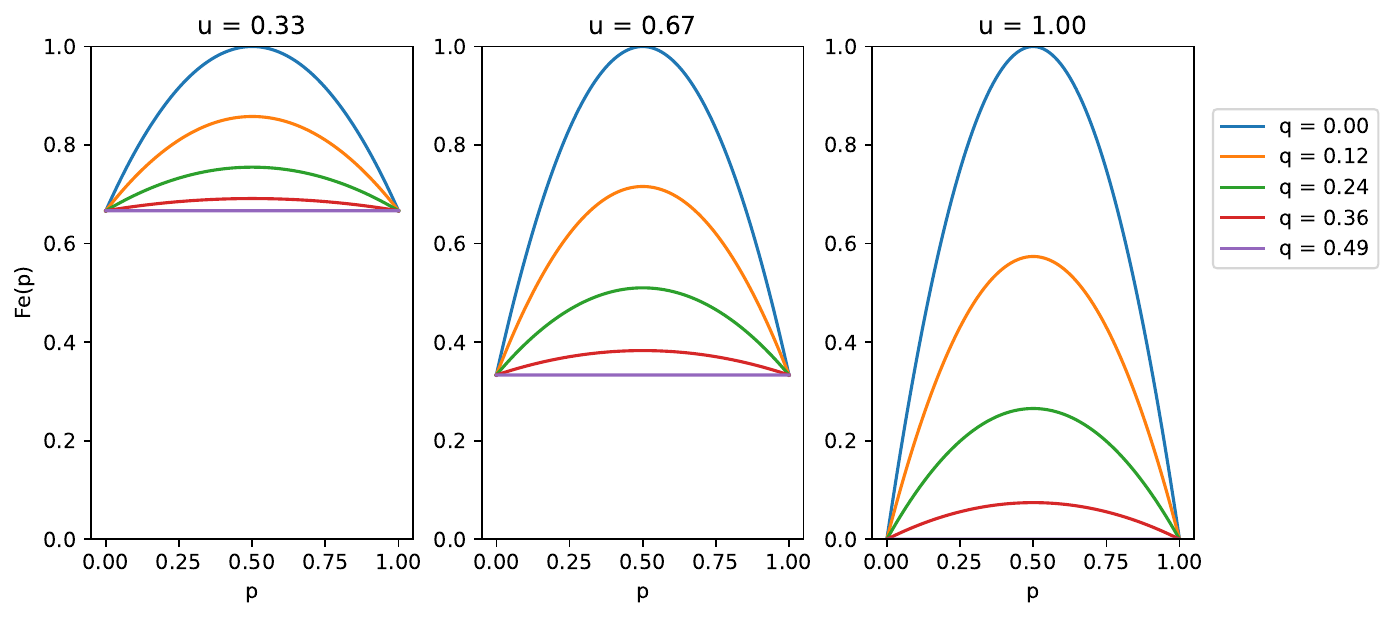}
\caption{Entanglement fidelity $F_e(p)$ for the random Pauli-$X$ channel, with fixed $u$ and $q$.}
    \label{fe-two-plt:pauli-x-2D}
    \centering
\end{figure}

Once $u$ and $q$ are fixed, our goal is to determine which value of $p$ maximizes entanglement preservation. Figure~\ref{fe-two-plt:pauli-x-2D} indicates that the entanglement fidelity is maximized near $p=0.5$ whenever $q \neq 1/2$. For $q=1/2$, the entanglement fidelity is independent of $p$, and hence all values of $p$ are equally optimal.

When $p=0.5$, the two letters become orthogonal and the message effectively reduces to a classical one. Nevertheless, subject to any constraints on state preparation, one can retain the relevant nonclassical features of the message while improving robustness by choosing $p$ as close to $0.5$ as allowed.

The choice of $p$ is also influenced by the probability $q$ of transmitting one of the two letters. In particular, as $q$ approaches $0.5$, the entanglement fidelity becomes less sensitive to the precise value of $p$, thereby allowing a wider range of admissible $p$ values without significantly degrading performance.

\subsubsection{Dephasing Channel}\label{fe-two-sec:deph-an}
From Table~\ref{fe-two-tab:channels-fe}, the \emph{entanglement fidelity} of the dephasing channel is given by
\begin{equation}
F_e(\rho,\mathcal{N}) \;=\; 1 + u\bigl[4p(p-1)\bigr].
\label{eq:fe_dephasing}
\end{equation}

\begin{figure}[t!]
    \includegraphics[width=0.5\textwidth]{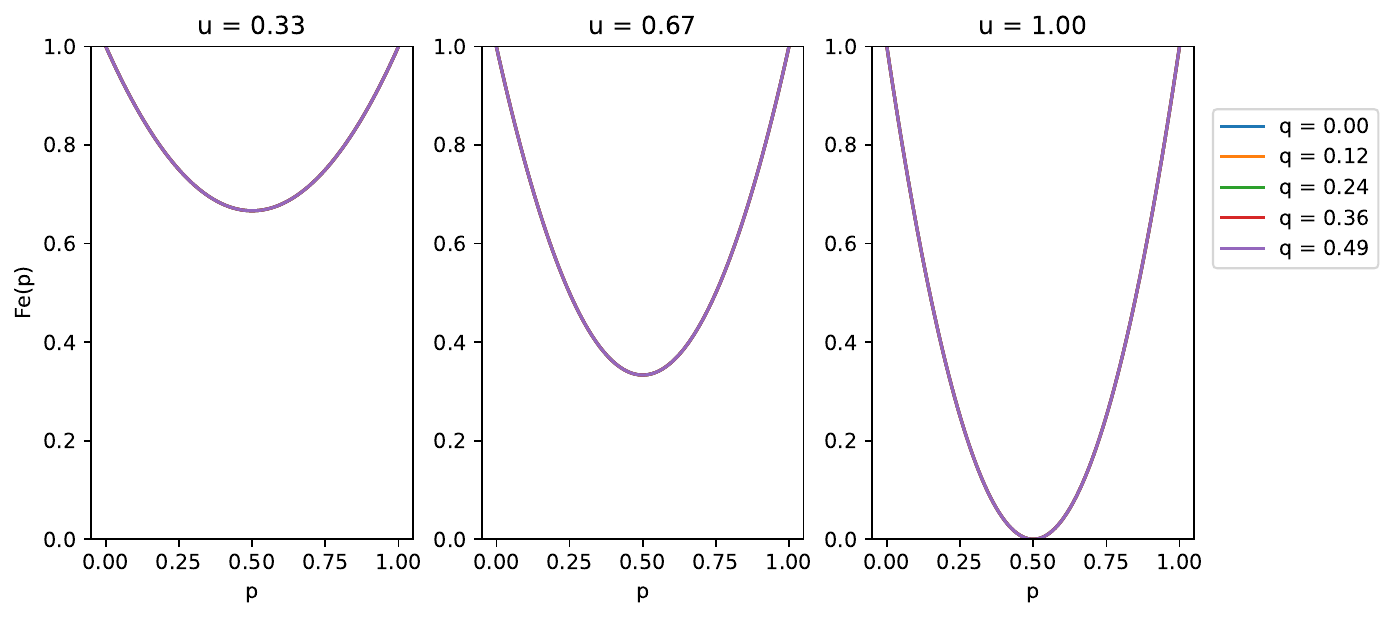}
\caption{Entanglement fidelity $F_e(p)$ for the dephasing channel, with fixed $u$ and $q$.}
    \label{fe-two-plt:deph-2D}
    \centering
\end{figure}

In this case, $F_e$ depends only on $u$ and $p$. Once $u$ is fixed, the entanglement fidelity is independent of $q$ and is a convex quadratic function of $p$ (see Fig.~\ref{fe-two-plt:deph-2D}). Consequently, it attains a unique minimum at $p = 0.5$.

In contrast to the random Pauli-$X$ channel (Section~\ref{fe-two-sec:pauli-x-an}), the largest values of $F_e$ are attained as $p$ approaches the endpoints of the interval. However, these limiting regimes correspond to degenerate alphabets in which the two-letter states coincide and therefore cannot convey information. In practice, one should therefore choose $p$ as close to $0$ or $1$ as the application allows to maximize entanglement preservation while avoiding complete degeneracy.

Although such choices may be infeasible in practice, in the limiting regimes $p \to 0$ or $p \to 1$ the entanglement fidelity approaches unity, that is, one can theoretically obtain $F_e$ arbitrarily close to $1$.

\subsubsection{Depolarizing Channel}\label{fe-two-sec:depol-an}
From Table~\ref{fe-two-tab:channels-fe}, the \emph{entanglement fidelity} of the depolarizing channel is given by

\begin{equation}
F_e(\rho,\mathcal{N})
= 1 + u\left[\,(2q-1)^2\,p(1-p) + \frac{(2p-1)^2}{4} -\frac{3}{4}\,\right].
\label{eq:fe_depolarizing}
\end{equation}

\begin{figure}[!tbh]
    \includegraphics[width=\linewidth]{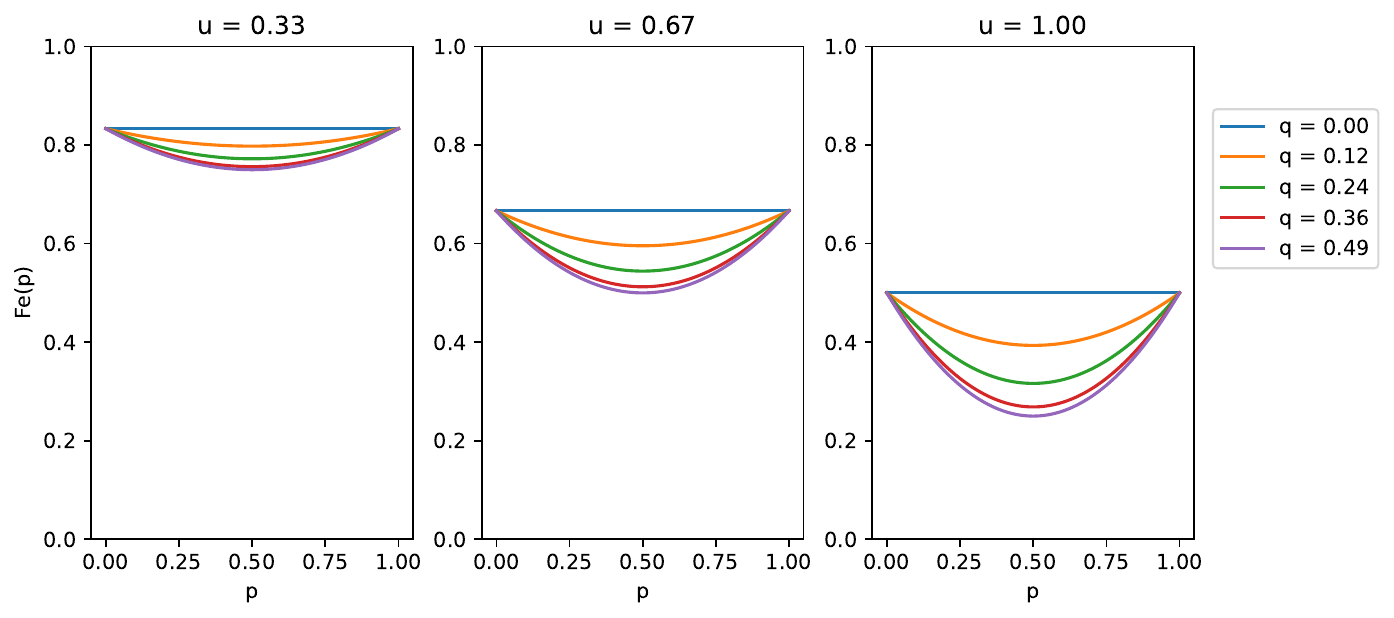}
    \caption{Entanglement fidelity $F_e(p)$ for the depolarizing channel, with fixed $u$ and $q$.}
    \label{fe-two-plt:depo-2D}
    \centering
\end{figure}

For fixed values of $u$ and $q$, $F_e$ is a convex quadratic function of $p$ (see Fig.~\ref{fe-two-plt:depo-2D}), with the limiting case $q \in \{0, 1\}$ yielding a constant function. As in the dephasing case (Section~\ref{fe-two-sec:deph-an}), the minimum is attained at $p=0.5$, while the largest values occur as $p$ approaches the endpoints of its admissible range.

A key difference between the two channels concerns the maximum achievable \emph{entanglement fidelity}. For the depolarizing channel, if $u \neq 0$, then no choice of $p$ can yield $F_e = 1$; perfect preservation is possible only in the noiseless case $u = 0$.

On the other hand, unlike the random Pauli-$X$ case discussed in Section~\ref{fe-two-sec:pauli-x-an}, the depolarizing channel does not completely destroy the entanglement even at $u = 1$: for all $p$ and $q$, the entanglement fidelity remains bounded away from zero, that is, $F_e$ admits a strictly positive minimum.

As suggested by the operational interpretation of the channel, larger values of $u$ lead to smaller achievable \emph{entanglement fidelity}. Indeed, as $u \to 1$, the contribution of the maximally mixed state $I/2$ to the channel output increases, which in turn reduces the maximum attainable $F_e$.

\subsubsection{Amplitude-damping Channel}\label{fe-two-sec:ad-an}
From Table~\ref{fe-two-tab:channels-fe}, the \emph{entanglement fidelity} of the amplitude-damping channel is given by
\begin{eqnarray}
F_e(\rho,\mathcal{N}) & = & p^2 + (1-\gamma)(1-p)^2 \nonumber \\
& + & p(1-p)\!\left[2\sqrt{1-\gamma} + \gamma(2q-1)^2\right].
\label{eq:fe_amplitude_damping}
\end{eqnarray}

\begin{figure}[!tbh]
    \includegraphics[width=0.5\textwidth]{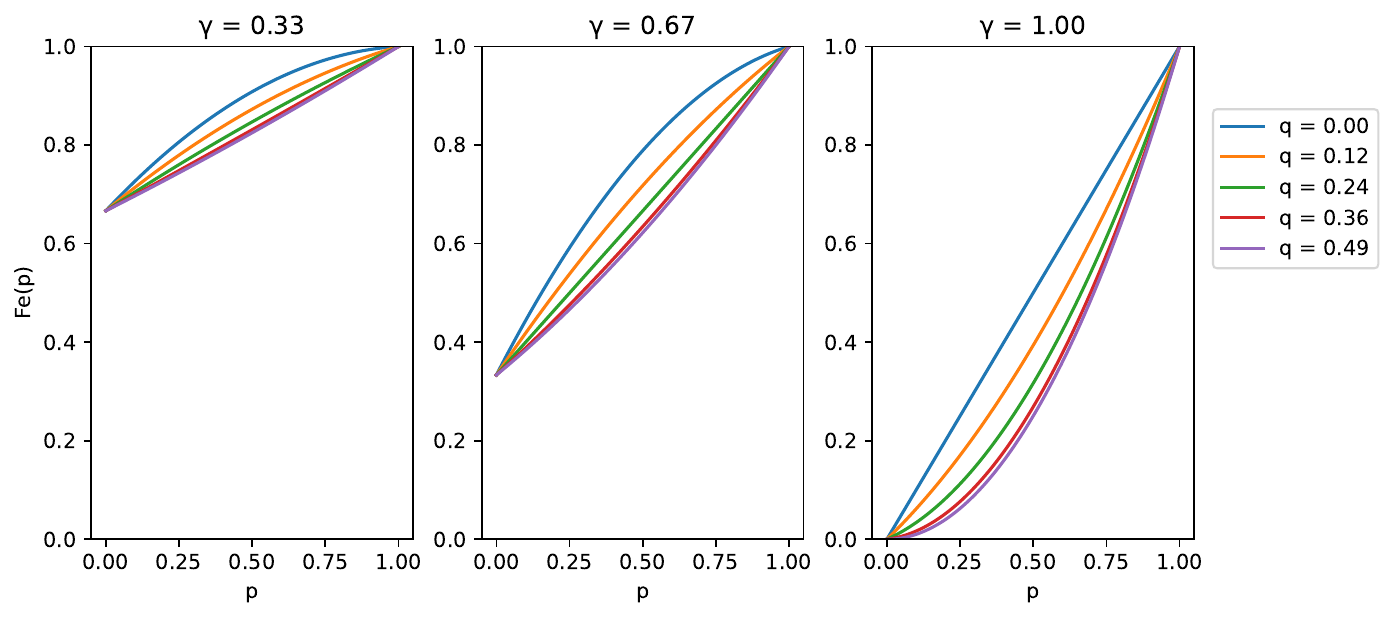}
\caption{Entanglement fidelity $F_e(p)$ for the amplitude-damping channel, with fixed $\gamma$ and $q$.}
    \label{fe-two-plt:ampdamp-2D}
    \centering
\end{figure}

Unlike in the previous cases, the function is not symmetric about $p = 0.5$ (see Fig.~\ref{fe-two-plt:ampdamp-2D}). Moreover, its convexity may depend on the value of $q$. Nevertheless, as $p$ approaches $1$, the entanglement fidelity approaches $1$ as well.

Although this limiting regime is not representative of practical applications, we conclude that, for an \emph{amplitude-damping} channel, it is advantageous to choose the letters in~\eqref{fe-two-eq:letters} so that $\ket{0}$ is the dominant component of the superposition.

\section{Ranking of Pauli Channels}\label{fe-two-sec:rank}
The Pauli channels introduced above share a common parametrization in terms of the noise strength $u$, while differing in the type of error they induce on the transmitted state. In Section~\ref{fe-two-sec:analysis}, we examined how, for a fixed channel, the letter parameter $p$ can be tuned to maximize the resulting \emph{entanglement fidelity}. Here, we address the complementary question: for a fixed source bias $q$ and across all admissible values of $p$, how do the Pauli channels compare in terms of entanglement preservation? Accordingly, we rank the three channels as functions of $p$, for a given $q$. Recalling the results in Table~\ref{fe-two-tab:channels-fe} and fixing $u$ and $q$, we define the following quantities:
\begin{itemize}
    \item Random Pauli-$X$ channel: $F_e^X(p) = 1 + u\left[4(2q - 1)^2\,p(1-p) - 1\right]$
    \item Dephasing channel: $F_e^Z(p) = 1 + 4u\,p(p-1)$
    \item Depolarizing channel: $F_e^D(p) = 1 + u\left[(2q - 1)^2\,p(1-p) + (2p-1)^2/4 - 3/4\right]$
\end{itemize}

To simplify the notation in the discussion, we set $k := (2q-1)^2$. Moreover, we assume $u \neq 0$, since the case $u = 0$ corresponds to the identity channel and is therefore irrelevant in realistic noisy implementations.

By solving the corresponding inequalities, we obtain the following result:
\begin{equation}
F_e^{X}(p) \ge F_e^{Z}(p),
\ \forall\, p \in \left[
\frac{1}{2} - \frac{1}{2}\sqrt{\frac{k}{1+k}},
\ \frac{1}{2} + \frac{1}{2}\sqrt{\frac{k}{1+k}}
\right].
\end{equation}

\begin{equation}
F_e^{D}(p) \ge F_e^{Z}(p),\ \forall\, p \in \left[
\frac{1}{2} - \frac{1}{2}\sqrt{\frac{1+k}{3+k}},
\ \frac{1}{2} + \frac{1}{2}\sqrt{\frac{1+k}{3+k}}
\right].
\end{equation}

When analyzing the inequality $F_e^{D}(p) \ge F_e^{X}(p)$, two parameter regimes must be distinguished. This distinction arises because real-valued solutions do not exist when $q \in \left(\frac{1}{2}-\frac{1}{2\sqrt{3}},\frac{1}{2}+\frac{1}{2\sqrt{3}}\right)$.

For $q \in \left(\frac{1}{2}-\frac{1}{2\sqrt{3}},\frac{1}{2}+\frac{1}{2\sqrt{3}}\right)$, we have
\begin{equation}
F_e^{D}(p) \ge F_e^{X}(p) \ \text{for all} \ p\in[0,1].
\end{equation}

Otherwise, one obtains
\begin{align*}
F_e^{D}(p) \ge F_e^{X}(p),
\end{align*}
\begin{align*}
\text{for all} \ p \in
\left[
0,\ \frac{1}{2} - \frac{1}{2} \sqrt{\frac{3k-1}{1+3k}}
\right]
\cup
\left[
\frac{1}{2} + \frac{1}{2}\sqrt{\frac{3k-1}{1+3k}},\ 1
\right].
\end{align*}

\begin{table}[htbp]
\caption{Ranking of the entanglement fidelity for the random Pauli-$X$ ($F_e^{X}$), dephasing ($F_e^{Z}$), and depolarizing ($F_e^{D}$) channels for
$q \in \left[0,\frac{1}{2}-\frac{1}{2\sqrt{3}}\right]\cup\left[\frac{1}{2}+\frac{1}{2\sqrt{3}},1\right]$.}
    \centering
\renewcommand{\arraystretch}{2.2}
\begin{tabular}{|m{4cm}|m{3.7cm}|}
\hline
\multicolumn{1}{|c|}{\textbf{$p$}} & \multicolumn{1}{c|}{\textbf{Ranking}} \\ \hline
$\left[0,\frac{1}{2}-\frac{1}{2}\sqrt{\frac{1+k}{3+k}}\right)$
& $F_e^Z(p) \ge F_e^D(p) \ge F_e^X(p)$ \\ \hline
$\left[\frac{1}{2}-\frac{1}{2}\sqrt{\frac{1+k}{3+k}},\frac{1}{2}-\frac{1}{2}\sqrt{\frac{k}{1+k}}\right)$
& $F_e^D(p) \ge F_e^Z(p) \ge F_e^X(p)$ \\ \hline
$\left[\frac{1}{2}-\frac{1}{2}\sqrt{\frac{k}{1+k}},\frac{1}{2}-\frac{1}{2}\sqrt{\frac{3k-1}{1+3k}}\right)$
& $F_e^D(p) \ge F_e^X(p) \ge F_e^Z(p)$ \\ \hline
$\left[\frac{1}{2}-\frac{1}{2}\sqrt{\frac{3k-1}{1+3k}},\frac{1}{2}+\frac{1}{2}\sqrt{\frac{3k-1}{1+3k}}\right)$
& $F_e^X(p) \ge F_e^D(p) \ge F_e^Z(p)$ \\ \hline
$\left[\frac{1}{2}+\frac{1}{2}\sqrt{\frac{3k-1}{1+3k}},\frac{1}{2}+\frac{1}{2}\sqrt{\frac{k}{1+k}}\right)$
& $F_e^D(p) \ge F_e^X(p) \ge F_e^Z(p)$ \\ \hline
$\left[\frac{1}{2}+\frac{1}{2}\sqrt{\frac{k}{1+k}},\frac{1}{2}+\frac{1}{2}\sqrt{\frac{1+k}{3+k}}\right)$
& $F_e^D(p) \ge F_e^Z(p) \ge F_e^X(p)$ \\ \hline
$\left[\frac{1}{2}+\frac{1}{2}\sqrt{\frac{1+k}{3+k}},1\right]$
& $F_e^Z(p) \ge F_e^D(p) \ge F_e^X(p)$ \\ \hline
\end{tabular}
\label{fe-two-tab:rank-q-union}
\end{table}

We have the following inequalities, which are useful for relating the corresponding $p$-intervals:
\begin{equation*}
\sqrt{\frac{3k-1}{1+3k}}
\;\le\;
\sqrt{\frac{k}{1+k}}
\;\le\;
\sqrt{\frac{1+k}{3+k}}, \quad \text{for all $k\geq 1/3$}
\end{equation*}
\begin{equation}
\text{or equivalently, for all } q \not\in \left(\frac{1}{2}-\frac{1}{2\sqrt{3}},\frac{1}{2}+\frac{1}{2\sqrt{3}}\right).
\end{equation}

For $k < 1/3$, only the rightmost inequality is needed, and it holds for all $k\in[0, 1]$.

\begin{table}[htbp]
   \caption{Ranking of the entanglement fidelity for the random Pauli-$X$ ($F_e^{X}$), dephasing ($F_e^{Z}$), and depolarizing ($F_e^{D}$) channels, for
    $q \in \left(\frac{1}{2}-\frac{1}{2\sqrt{3}},\frac{1}{2}+\frac{1}{2\sqrt{3}}\right)$.}
    \centering
    \renewcommand{\arraystretch}{2.2} 
    \begin{tabular}{|m{4cm}|m{3.7cm}|} 
    \hline
    \multicolumn{1}{|c|}{\textbf{$p$}} & \multicolumn{1}{c|}{\textbf{Ranking}} \\ \hline    
    $\left[0, \frac{1}{2} - \frac{1}{2}\sqrt{\frac{1+k}{3 + k}}\right)$ & $F_e^Z(p) \ge F_e^D(p) \ge F_e^X(p)$ \\ \hline
    $\left[\frac{1}{2} - \frac{1}{2}\sqrt{\frac{1+k}{3+k}}, \frac{1}{2} - \frac{1}{2}\sqrt{\frac{k}{1+k}}\right)$ & $F_e^D(p) \ge F_e^Z(p) \ge F_e^X(p)$ \\ \hline
    $\left[\frac{1}{2} - \frac{1}{2}\sqrt{\frac{k}{1+k}}, \frac{1}{2} + \frac{1}{2}\sqrt{\frac{k}{1+k}}\right)$ & $F_e^D(p) \ge F_e^X(p) \ge F_e^Z(p)$  \\ \hline
    $\left[\frac{1}{2} + \frac{1}{2}\sqrt{\frac{k}{1+k}}, \frac{1}{2} + \frac{1}{2}\sqrt{\frac{1+k}{3+k}}\right)$ & $F_e^D(p) \ge F_e^Z(p) \ge F_e^X(p)$  \\ \hline
    $\left[\frac{1}{2} + \frac{1}{2}\sqrt{\frac{1+k}{3+k}}, 1\right]$ & $F_e^Z(p) \ge F_e^D(p) \ge F_e^X(p)$  \\ \hline
    \end{tabular}
    \label{fe-two-tab:rank-q-intersect}
\end{table}


\subsection{Commonly Used States}
Instead of considering two-letter messages, many quantum communication protocols involve transmitting a single fixed state through a channel (see, for example,~\cite{BB84,E91,Superdense-Coding,Teleportation}). The results developed above can be specialized to this setting by considering the limiting cases $q = 0$ or $q = 1$, for which the source emits a single letter with unit probability.

For these values of $q$, the intervals reported in Tables~\ref{fe-two-tab:rank-q-union} and~\ref{fe-two-tab:rank-q-intersect} simplify, and the corresponding channel ranking is summarized in Table~\ref{fe-two-tab:rank-q-01}.

\begin{table}[htbp]
    \caption{Ranking of the entanglement fidelity for the random Pauli-$X$ ($F_e^{X}$), dephasing ($F_e^{Z}$), and depolarizing ($F_e^{D}$) channels for $q=0$ or $q=1$.}
    \centering
    \renewcommand{\arraystretch}{2.2}
    \begin{tabular}{|m{3.8cm}|m{4cm}|}
    \hline
    \multicolumn{1}{|c|}{\textbf{$p$}} & \multicolumn{1}{c|}{\textbf{Ranking}} \\ \hline
    $\left[0,\frac{1}{2}-\frac{1}{2\sqrt{2}}\right)$ & $F_e^Z(p) \ge F_e^D(p) \ge F_e^X(p)$ \\ \hline
    $\left[\frac{1}{2}-\frac{1}{2\sqrt{2}},\frac{1}{2}+\frac{1}{2\sqrt{2}}\right)$ & $F_e^X(p) \ge F_e^D(p) \ge F_e^Z(p)$ \\ \hline
    $\left[\frac{1}{2}+\frac{1}{2\sqrt{2}},1\right]$ & $F_e^Z(p) \ge F_e^D(p) \ge F_e^X(p)$ \\ \hline
    \end{tabular}
    \label{fe-two-tab:rank-q-01}
\end{table}

By varying $p$ and $q$, we obtain channel-performance comparisons for different source states. Table~\ref{fe-two-tab:rank-special-states} summarizes the resulting rankings for several commonly used qubit states.

\begin{table}[htbp]
\caption{Ranking of the entanglement fidelity for the random Pauli-$X$ ($F_e^{X}$), dephasing ($F_e^{Z}$), and depolarizing ($F_e^{D}$) channels for commonly used qubit states.}
    \centering
    \renewcommand{\arraystretch}{2.2} 
    \begin{tabular}{|m{0.4cm}|m{0.4cm}|m{2.7cm}|m{3.5cm}|} 
    \hline
    \multicolumn{1}{|c|}{\textbf{$q$}} & \multicolumn{1}{|c|}{\textbf{$p$}} & \multicolumn{1}{|c|}{\textbf{State}} & \multicolumn{1}{c|}{\textbf{Ranking}} \\ \hline    
    1  & 1 & $\ket{0}$ & $F_e^Z(p) \ge F_e^D(p) \ge F_e^X(p)$ \\ \hline
    1  & 0 & $\ket{1}$ & $F_e^Z(p) \ge F_e^D(p) \ge F_e^X(p)$ \\ \hline
    1  & $\frac{1}{2}$ & $\ket{+} = \frac{1}{\sqrt{2}}\left(\ket{0} + \ket{1}\right)$ & $F_e^X(p) \ge F_e^D(p) \ge F_e^Z(p)$ \\ \hline
    0  & $\frac{1}{2}$ & $\ket{-} = \frac{1}{\sqrt{2}}\left(\ket{0} - \ket{1}\right)$ & $F_e^X(p) \ge F_e^D(p) \ge F_e^Z(p)$ \\ \hline
    \end{tabular}
    \label{fe-two-tab:rank-special-states}
\end{table}

In the study of composite systems, an important and ubiquitous family of states is the Bell basis:
\begin{itemize}
    \item \( \ket{\Phi^{+}} = \frac{1}{\sqrt{2}}\bigl(\ket{00} + \ket{11}\bigr) \)
    \item \( \ket{\Phi^{-}} = \frac{1}{\sqrt{2}}\bigl(\ket{00} - \ket{11}\bigr) \)
    \item \( \ket{\Psi^{+}} = \frac{1}{\sqrt{2}}\bigl(\ket{01} + \ket{10}\bigr) \)
    \item \( \ket{\Psi^{-}} = \frac{1}{\sqrt{2}}\bigl(\ket{01} - \ket{10}\bigr) \)
\end{itemize}

If the composite system formed by the letter state and the reference system is prepared in any of these states, then the entanglement between the two subsystems is maximal. The corresponding density operators are
\begin{center}
$\ket{\Phi^{+}}\!\bra{\Phi^{+}},\ \ket{\Phi^{-}}\!\bra{\Phi^{-}},\ \ket{\Psi^{+}}\!\bra{\Psi^{+}},\ \ket{\Psi^{-}}\!\bra{\Psi^{-}}.$
\end{center}

By tracing out the reference system, we find that each Bell state is a purification of the same reduced density operator:
\begin{equation}\label{fe-two-eq:rho-bell-states}
\rho
= \mathrm{Tr}_{R}\!\left(\ket{\beta}\!\bra{\beta}\right)
= \frac{1}{2}\bigl(\ket{0}\!\bra{0}+\ket{1}\!\bra{1}\bigr)
= \frac{1}{2}\bigl(\ket{+}\!\bra{+}+\ket{-}\!\bra{-}\bigr),
\end{equation}
where $\ket{\beta}\in\{\ket{\Phi^{\pm}},\ket{\Psi^{\pm}}\}$ and $\ket{\pm}=(\ket{0}\pm\ket{1})/\sqrt{2}$.

Moreover, by setting $q=\tfrac{1}{2}$ and $p=\tfrac{1}{2}$, the density operator in~\eqref{fe-two-eq:rho-bell-states} can be written as an equal mixture of the letter states $\ket{\psi^{+}}$ and $\ket{\psi^{-}}$:
\begin{equation}
\rho \;=\; \frac{1}{2}\Bigl(\ket{\psi^{+}}\!\bra{\psi^{+}} \;+\; \ket{\psi^{-}}\!\bra{\psi^{-}}\Bigr).
\end{equation}

In this setting, the source effectively emits a two-letter message. Moreover, within our two-letter source model, the Bell-state reduced density operator $\rho=I/2$ is obtained precisely for $p=1/2$ and $q=1/2$. It follows that $F_e^{X}=F_e^{Z}=1-u$, whereas $F_e^{D}=1-u/2$. Hence, if the initial joint state of the letter and reference systems is a Bell state, the Pauli-channel ranking is
\begin{center}
$F_e^{D} \ge F_e^{X} = F_e^{Z}$.
\end{center}

\section{Conclusion}
\label{sec:conclusion}
We studied entanglement fidelity as a quantitative measure of how well a noisy quantum channel preserves not only an input state but also its correlations with an external reference. Using the Kraus-operator formulation, we developed a channel-agnostic procedure for calculating $F_e(\rho,\mathcal{E})$ for any CPTP map with a finite Kraus representation and derived explicit expressions for several standard channels, including the random Pauli-$X$, dephasing, depolarizing, Werner-Holevo, generalized Pauli, and amplitude-damping channels.

We then applied these formulas to a communication setting in which a two-letter parametric message is transmitted. This framework makes explicit how $F_e$ depends jointly on channel and message parameters, thereby enabling direct comparisons across channels and letter choices. In particular, we used the resulting expressions to rank Pauli-type channels for relevant families of message states, including commonly used qubit states and the maximally mixed reduced state associated with Bell-state purifications. These results provide practical guidance for selecting encodings that improve entanglement preservation under a given noise model.

Several extensions follow naturally. Our approach can be extended directly to other channels and higher-dimensional systems, including structured noise models whenever an operator-sum representation is available. It would also be interesting to combine these results with explicit optimization under state-preparation constraints, larger alphabets, and coding or recovery schemes.

\section*{Acknowledgment}
This work is partially supported by the European Research Council (ERC) under the EU's Horizon 2020 research and innovation programme (Grant agreement No. 101003431). 



\end{document}